\documentclass[namedreferences]{solarphysics}

\usepackage[hyperref,optionalrh,showbiblabels]{spr-sola-addons} 
\usepackage{dcolumn}
\usepackage{graphicx}        
\usepackage{amssymb} 
\usepackage{amstext}        
\usepackage{color}           
\usepackage{breakurl}        
\usepackage{tabularx}
\usepackage{multirow}

\newcommand{\aap}{    {\it Astron. Astrophys.}}

\newcommand{\apj}{    {\it Astrophys. J.}}
\newcommand{\apjl}{   {\it Astrophys. J. Lett.}}

\newcommand{\nat}{    {\it Nature}}

\newcommand{\solphys}{{\it Solar Phys.}}
 
\newcommand{\ssr}{    {\it Space Sci. Rev.}} 
\chardef\us=`\_

\begin{document}

\begin{article}
\begin{opening}

\title{Identification of Pre-flare Processes and Their Possible Role in Driving a Large-scale Flux Rope Eruption with Complex M-class Flare in the Active Region NOAA 12371}

\author[addressref={aff1,aff2},corref,email={prabir@prl.res.in}]{\inits{P. K.}\fnm{Prabir K.}~\lnm{Mitra}\orcid{https://orcid.org/0000-0002-0341-7886}}
\author[addressref=aff1]{\inits{B.J.}\fnm{Bhuwan}~\lnm{Joshi}\orcid{https://orcid.org/0000-0001-5042-2170}}
\author[addressref=aff3]{\inits{A.P.}\fnm{Avijeet}~\lnm{Prasad}\orcid{https://orcid.org/0000-0003-0819-464X}}

\address[id=aff1]{Udaipur Solar Observatory, Physical Research Laboratory, Udaipur 313 001, India}
\address[id=aff2]{Department of Physics, Gujarat University, Ahmedabad 380 009, India}
\address[id=aff3]{Center for Space Plasma and Aeronomic Research, The University of Alabama in Huntsville, Huntsville, Alabama 35899, USA}

\runningauthor{P. K. Mitra \textit{et al.}}
\runningtitle{Pre-flare Processes and Flux Rope Eruption}

\begin{abstract}
In this article, we study the origin of precursor flare activity and investigate its role towards triggering the eruption of a flux rope which resulted into a dual-peak M-class flare (SOL2015-06-21T02:36) in the active region NOAA 12371. The flare evolved in two distinct phases with peak flux levels of M2.1 and M2.6 at an interval of $\approx$54 min. The active region exhibited striking moving magnetic features (MMFs) along with sunspot rotation. Non-linear force free field (NLFFF) modelling of the active region corona reveals a magnetic flux rope along the polarity inversion line in the trailing sunspot group which is observationally manifested by the co-spatial structures of an active region filament and a hot channel identified in the 304 and 94 \AA\ images, respectively, from the \textit{Atmospheric Imaging Assembly} (AIA). The active region underwent a prolonged phase of flux enhancement followed by a relatively shorter period of flux cancellation prior to the onset of the flare which led to the build up and activation of the flux rope. Extreme ultra-violet (EUV) images reveal localised and structured pre-flare emission, from the region of MMFs, adjacent to the location of the main flare. Our analysis reveals strong, localised regions of photospheric currents of opposite polarities at the precursor location, thereby making the region susceptible to small-scale magnetic reconnection. Precursor reconnection activity from this location most likely induced a slipping reconnetion towards the northern leg of the hot channel which led to the destabilization of the flux rope. The application of magnetic virial theorem suggests that there was an overall growth of magnetic free energy in the active region during the prolonged pre-flare phase which decayed rapidly after the hot channel eruption and its successful transformation into a halo coronal mass ejection (CME).
\end{abstract}

\end{opening}

%
\section{Introduction} \label{intro}
Solar flares belong to the most spectacular phenomena occurring in the solar system. During a flare, catastrophic energy release of the order as high as 10$^{27}$--10$^{32}$ erg occurs in the solar atmosphere within tens of minutes. The released energy manifests its signatures in the entire electromagnetic spectrum in the form of heat and particle acceleration \citep[see reviews by][]{Fletcher2011, Benz2017}. Major flares are often associated with large-scale eruption of plasma from the solar corona known as coronal mass ejection (CME). Earth-directed CMEs are known to cause geomagnetic storms and other hazardous effects at the near-Earth environment. The precise understanding of the magnetic configuration of the flare producing active regions (ARs) during the pre-flare phase and its role in triggering the large-scale eruptions are among the most critical topics studied in the solar physics community \citep[see \textit{e.g.},][]{Mitra2019}. 

The ``standard flare model'', also known as CSHKP model \citep{Carmichael1964, Sturrock1966, Hirayama1974, Kopp1976}, recognizes the presence of magnetic flux rope (MFR) in the AR corona as the prerequisite for the initiation of eruptive flares. An MFR is recognised as sets of magnetic field lines which are twisted around its central axis more than once \citep{Gibson2006}. These complex structures are believed to be formed as a result of flux cancellation over the polarity inversion line (PIL) through photospheric shearing and converging motions \citep{van1989}; however, the exact mechanism for flux rope formation is still unclear and debatable. Observationally, MFRs have been identified in the form of different solar features, such as, filaments, prominences, filament channels, hot coronal channels, and coronal sigmoids. Filaments are threadlike structures which are observed as dark long narrow features in the chromospheric images of the Sun \citep{Zirin1988, Martin1998}. When these structures are observed over the limb, they appear brighter than the background sky and are called prominences \citep{Tandberg1995, parenti2014}. Filament channels are voids without plages or chromospheric fine structures such as spicules or fibrils \textit{etc.} \citep{Martres1966, Gaizauskas1997}. These are long, narrow, extended structures situated over the PIL where filaments or prominences are formed \citep{Engvold1997}.

While filaments and filament channels are observed in the absorption lines of choromospheric images, sigmoids and hot channels are observed in the emission line features of the solar corona. Sigmoids are ``S'' (or inverted ``S'') shaped structures that are observed in the soft X-ray (SXR) and the extreme ultra-violet (EUV) images of the Sun \citep{Rust1996, Manoharan1996}. Hot channels are coherent structures observed in the high temperature pass-band EUV images of the solar corona \citep{Zhang2012, Cheng2013}. These are often found in association with coronal sigmoids \citep[see \textit{e.g.},][]{Cheng2014b, Joshi2017, Mitra2018}. Their frequent co-existence with filaments confirms that filaments and hot channels are different observational manifestations of MFRs lying in the chromospheric and coronal heights, respectively \citep{Cheng2014}. However, the most important feature of hot channels and coronal sigmoids is their frequent association with CMEs which has been suggested by several case studies and statistical surveys \citep[see \textit{e.g.},][]{Nindos2015}.

The temporal evolution of a typical eruptive flare can be summarized in three phases: pre-flare/precursor phase, impulsive phase, and gradual phase. While the processes occurring during the impulsive and gradual phases are broadly explained by the CSHKP model \citep{Shibata1996}, the pre-flare phase is still ill-understood. Pre-flare activities are considered to be important in order to understand the physical conditions that lead to flares and associated eruptions \citep[see \textit{e.g.},][]{Farnik1996, Chifor2006, Joshi2011}. Although, it is well understood that a significant fraction of all the major flares are associated with pre-flare events, the causal relation between them requires investigation in detail through multi-wavelength case studies. Thanks to the high resolution and high cadence observations of the \textit{Atmospheric Imaging Assembly} \citep[AIA;][]{Lemen2012} and the \textit{Helioseismic and Magnetic Imager} \citep[HMI;][]{Schou2012} on board the \textit{Solar Dynamics Observatory} \citep[SDO;][]{Pesnell2012}, in the recent years, there has been a progress in understanding the short lived pre-flare events and precursor emission.

In this article we present a detailed multi-wavelength analysis of the pre-flare processes associated with a dual-peak M-class flare on 21 June 2015. The reported events occurred in the AR NOAA 12371 which was among the prominent ARs of the solar cycle 24. The M-class flares produced by this AR on 21 and 22 June 2015 have been subjected to a number of studies \citep[see \textit{e.g.},][]{Manoharan2016, Cheng2016b, Jing2017, Vemareddy2017, Piersanti2017, Bi2017, Lee2017, Wang2018, Lee2018, Kuroda2018, Joshi2018, Gopalswamy2018, Liu2019}. Our rigorous analysis aims towards providing clear understanding of the pre-flare energy release processes and the role of the pre-flare activity in triggering the flux rope eruption during the complex M-class flare that displayed characteristics of a long duration event (LDE) of energy release. Section \ref{data} provides a detailed account of the observational data and analysis techniques used in this article. In Section \ref{ar}, we discuss the evolution of the AR NOAA 12371 in detail. The results obtained on the basis of multi-wavelength analysis and photospheric measurements are discussed in Section \ref{obs}. Non-linear force free field (NLFFF) extrapolation results and evolution of magnetic free energy in the AR are provided in Section \ref{sec_nlfff}. We discuss and interpret our results in Section \ref{discus}.

\section{Observational Data and Methods} \label{data}
Solar observation in EUV wavelengths were obtained from the \textit{Atmospheric Imaging Assembly} \citep[AIA;][]{Lemen2012} on board the \textit{Solar Dynamics Observatory} \citep[SDO;][]{Pesnell2012}. Among the seven EUV filters of AIA (94 \AA , 131 \AA , 171 \AA , 193 \AA , 211 \AA , 304 \AA , and 335 \AA ), we have focussed on the 4096$\times$4096 pixel full-disk solar images in the 94 and 304 \AA\ channels at a spatial resolution of $0.''$6 pixel$^{-1}$ and temporal cadence of 12 s.

For studying the photospheric structures and their evolution, we have used intensity and magnetogram images taken by the \textit{Helioseismic and Magnetic Imager} \citep[HMI;][]{Schou2012} on board SDO. HMI produces full-disk line of sight (LOS) intensity (continuum) and magnetogram images of 4096$\times$4096 pixels at a spatial resolution of $0.''$5 pixel$^{-1}$ and 45 s temporal cadence while the vector magnetograms are produced with a temporal cadence of 720 s.

Coronal magnetic field extrapolation has been carried out by employing the optimization based non-linear force free field (NLFFF) extrapolation method developed by \citet{Wiegelmann2010, Wiegelmann2012}, using photospheric vector magnetograms from the ``hmi.sharp\_cea\_720s'' series of HMI/SDO at a spatial resolution of $1.''$0 pixel$^{-1}$ as boundary conditions. Extrapolations have been done in a Cartesian volume of 474$\times$226$\times$226 pixels which translates to a physical volume with dimensions $\approx$344$\times$164$\times$164 Mm. Using NLFFF extrapolation results, we calculated the squashing factor (\textit{Q}) in the extrapolation-volume by employing the code introduced by \citet{Liu2016}. For visualizing the extrapolated coronal field lines, we have used Visualization and Analysis Platform for Ocean, Atmosphere, and Solar Researchers \citep[VAPOR;][]{Clyne2007} software.

X-ray observation during the flare was provided by the \textit{Reuven Ramaty High Energy Solar Spectroscopic Imager} \citep[RHESSI;][]{Lin2002}. RHESSI observed the full Sun with an unprecedented spatial resolution (as fine as $\approx2.''3$) and energy resolution (1--5 keV) in the energy range 3 keV--17 MeV. For imaging of the X-ray sources using RHESSI observation, we have used the CLEAN algorithm \citep{Hurford2002} with the natural weighting scheme for front detector segments 2--9 (excluding 7).

\section{Magnetic Structure of the Active Region NOAA 12371} \label{ar}

\subsection{Morphology and Build-up of Magnetic Complexity} \label{morp}
\begin{figure}
\includegraphics[width=1\textwidth]{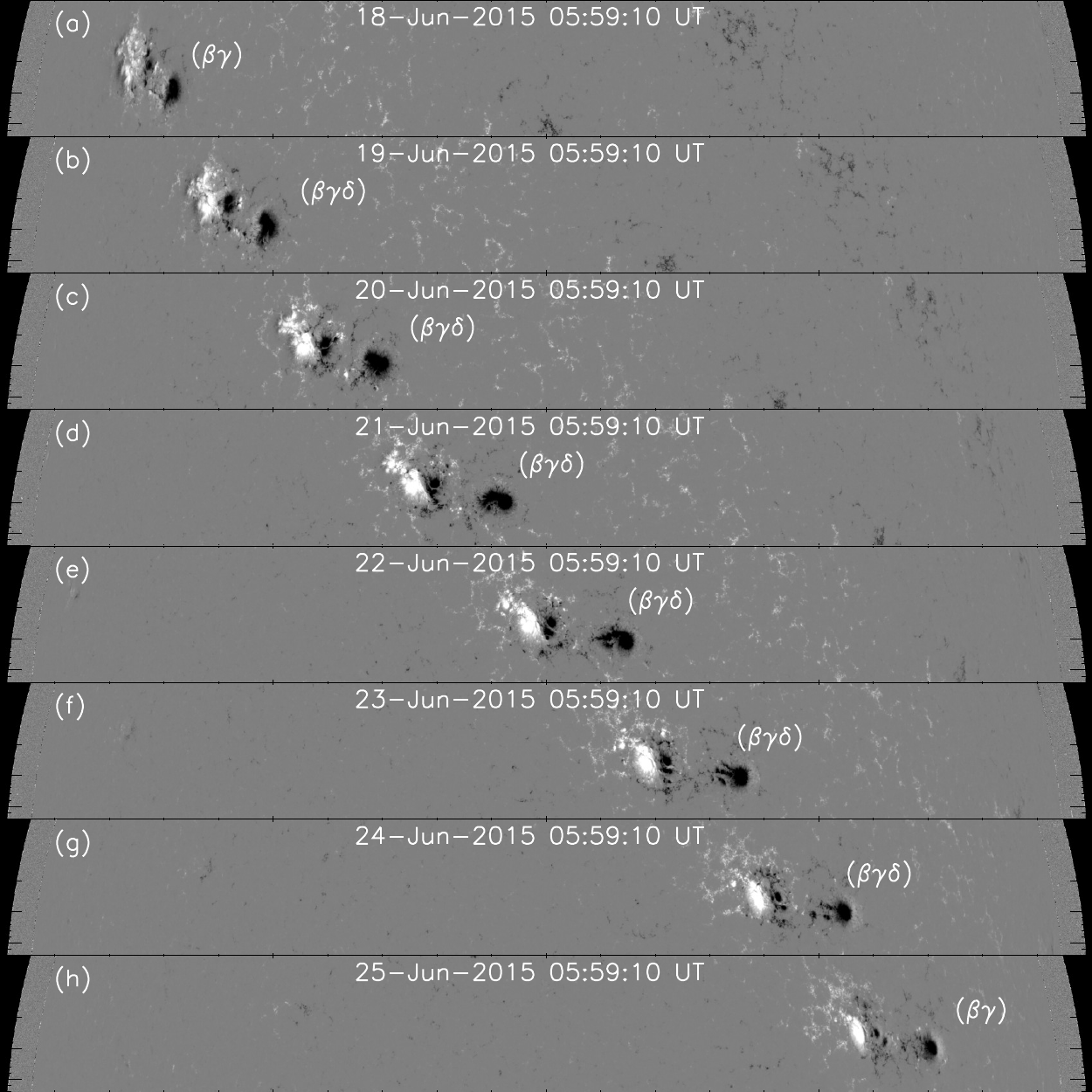}
\caption{Line-of-sight (LOS) magnetograms showing the synoptic overview of the photospheric magnetic structure of the AR NOAA 12371 during 18--25 June 2015. Magnetic configuration of the active region on each day is annotated in the corresponding panel.}
\label{ar_ev}
\end{figure}

The AR NOAA 12371 appeared on the eastern limb of the Sun during the last hours of 15 June 2015 as a moderately complex $\beta$-type sunspot. Flaring activity from the AR was noted to start from 16 June 2015. Interestingly, despite producing numerous B and C class flares alongside few M-class ones, it did not produce any X-class flare. By 18 June 2015, it gradually became a more complex $\beta\gamma$-type sunspot (Figure \ref{ar_ev}a) and produced its first M-class flare on the same day. The AR evolved into the most complex $\beta\gamma\delta$-type (Figure \ref{ar_ev}b) on the next day, \textit{i.e.} 19 June 2015. The highest flaring activity from the AR was observed during 20--22 June 2015. In this duration, it produced 3 M-class flares along with many C and B-class flares. The AR started to decay after 22 June 2015. In Table \ref{table1}, we summarise various evolutionary aspects of NOAA 12371 during its disk passage during 18--25 June 2015 which are collected from the NOAA Solar Region Summary (SRS) reports\footnote{see https://www.swpc.noaa.gov/products/solar-region-summary.}. The magnetic configuration of the AR reduced to $\beta\gamma$-type on 25 June 2015 (Figure \ref{ar_ev}h) and even further to $\beta$-type on 29 June 2015 just before it went on to the far side of the Sun from the western limb. During its lifetime on the visible hemisphere of the Sun, it produced a total of 5 M-class flares which are also listed in Table \ref{table1}. Notably, the AR produced the largest flare of class M7.9 on 25 June 2015 when it had entered into the declining phase. However, in terms of space weather manifestations, the most interesting eruptive flare from the AR occurred on 21 June 2015. This event followed the characteristics of a long duration flare with peak GOES flux reaching a level of M2.6 and a fast CME\footnote{see https://cdaw.gsfc.nasa.gov/CME\_list/index.html.} with linear speed of 1366 km s$^{-1}$.

\begin{table}
\caption{Evolution of the AR NOAA 12371 during its peak activity period and summary of the major flares produced by it.}
\label{table1}
\begin{tabular}{cccccc}
\hline
\hline
Sr.&Date&Heliographic&AR&Area (Millionth&GOES flare class\\
No.&&co-ordinates&configuration&of hemisphere)&(peak time (UT))\\
\hline
1&2015 June 18&$\approx$N12E53&$\beta\gamma$&$\approx$520&M3.0 (17:36)\\
2&2015 June 19&$\approx$N12E39&$\beta\gamma\delta$&$\approx$810&---\\
3&2015 June 20&$\approx$N13E27&$\beta\gamma\delta$&$\approx$1020&M1.0 (06:48)\\
4&2015 June 21&$\approx$N13E14&$\beta\gamma\delta$&$\approx$1120&M2.6 (02:36)\\
5&2015 June 22&$\approx$N13W00&$\beta\gamma\delta$&$\approx$1180&M6.5 (18:23)\\
6&2015 June 23&$\approx$N13W13&$\beta\gamma\delta$&$\approx$1070&---\\
7&2015 June 24&$\approx$N12W28&$\beta\gamma\delta$&$\approx$950&---\\
8&2015 June 25&$\approx$N11W40&$\beta\gamma$&$\approx$740&M7.9 (08:16)\\
\hline
\end{tabular}
\end{table}

\subsection{Evolution of Photospheric Magnetic Flux} \label{pfmg}
\begin{figure}
\includegraphics[width=0.85\textwidth]{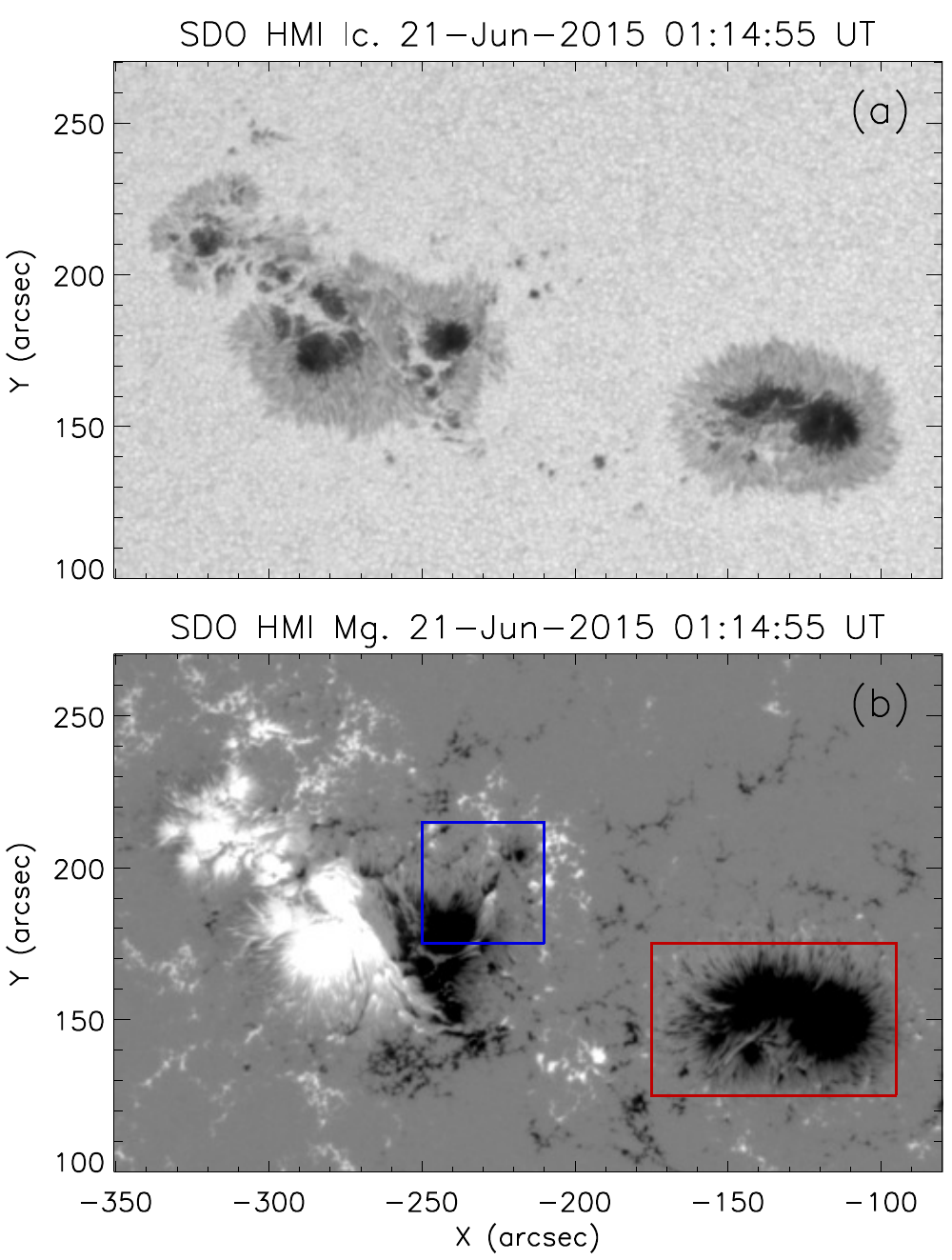}
\caption{(\textbf{a}): a white light image of the AR during the pre-flare phase of the M-class flare on 21 June 2015. (\textbf{b}): a co-temporal LOS magnetogram. The blue and red boxes indicate two regions that showed striking photospheric changes. An animation of this figure is provided in the supplementary materials.}
\label{ar_phot}
\end{figure}

To have a comprehensive understanding of the photospheric magnetic activities during the prolonged phase of energy build up and subsequent flaring activity on 21 June 2015, we have thoroughly examined large as well as small-scale changes in  the LOS magnetograms. During this period, the AR was consisted of two major sunspot groups (Figure \ref{ar_phot}a). A comparison of the white light image of the AR with a co-temporal magnetogram (\textit{cf.} Figures \ref{ar_phot}a and b) reveals that almost all of the leading sunspot group of the AR was consisted of negative polarity while the trailing sunspot group was of mixed polarities forming $\delta$-type configuration. In this manner, the overall photospheric configuration of magnetic polarities of the AR made it a $\beta\gamma\delta$ sunspot region. 

\begin{figure}
\includegraphics[width=1\textwidth]{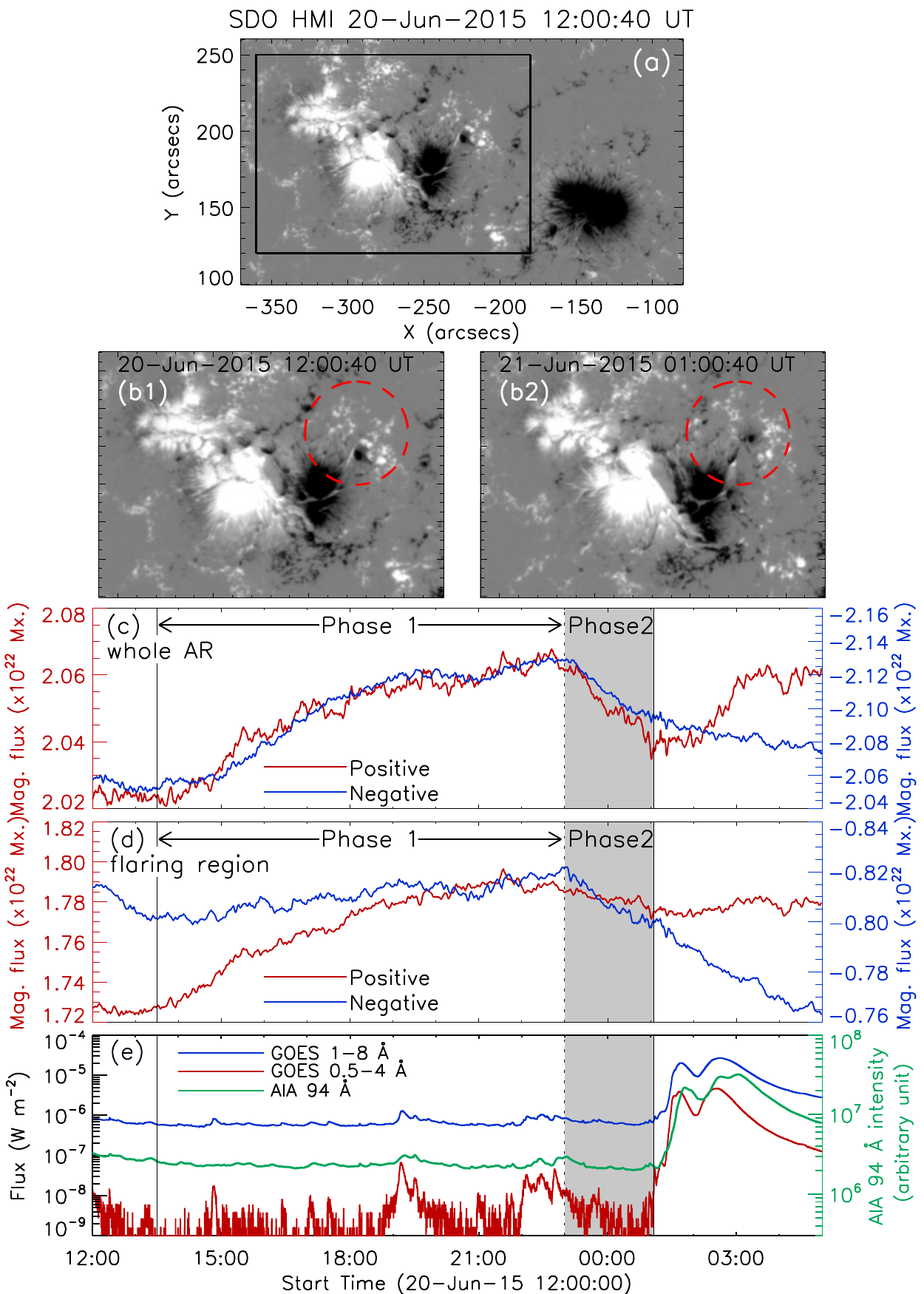}
\caption{(\textbf{a}): HMI LOS magnetogram of AR 12371. We outline the flaring region in the AR by the black box. (\textbf{b1}) and (\textbf{b2}): the flaring region as indicated by the black box in (\textbf{a}), at two different times with a time-gap of $\approx$23 hours. The red dashed circles in these two panels mark the region with striking moving magnetic features (MMFs). (\textbf{c}): evolution of magnetic flux in the whole AR from 20 June 2015 12:00 UT to 21 June 2015 05:00 UT. (\textbf{d}): evolution of magnetic flux in the flaring region (within the box in (\textbf{a})). In (\textbf{e}), we plot GOES SXR light curves in both the channels as well as the AIA 94 \AA\ intensity profile. In (\textbf{c})--(\textbf{e}), the interval marked as ``phase 1'' depicts flux emergence of both polarities in the AR as well as flaring region. The ``phase 1'' was followed by a period of flux cancellation of both polarities (``phase 2''; the shaded region) which lasts up to the initiation of the M-class flare on 21 June 2015.}
\label{hmi_lightcurve}
\end{figure}

In Figure \ref{hmi_lightcurve}, we plot the evolution of the LOS magnetic flux through the whole AR as well as only from the trailing sunspot region from 20 June 2015 12:00 UT to 21 June 2015 05:00 UT. The trailing sunspot group has been shown within the box in Figure \ref{hmi_lightcurve}a. Notably, the eruptive flare under investigation was triggered from this region. The evolution of the trailing sunspot group of the AR can be inferred by comparing Figures \ref{hmi_lightcurve}b1 and \ref{hmi_lightcurve}b2 which present enlarged view of the selected region at two different times. The area marked by dashed circles shows rapidly evolving magnetic elements and moving magnetic features (MMFs) which are discussed in the next subsection. From Figure \ref{hmi_lightcurve}c, we readily find that both the positive and negative flux from the AR increased during $\approx$13:30 UT--23:00 UT on 20 June 2015 and thereafter decreased till the onset of the flare at $\approx$01:05 UT on 21 June 2015. Based on the flux variations, we have divided the whole interval in two phases: ``phase 1''; when the active region displayed flux enhancement in both the polarities, and ``phase 2''; when flux of both polarities decayed. Evolution of magnetic flux in the trailing sunspot group, (\textit{i.e.} the flaring region) displayed similar variation as the entire AR (Figure \ref{hmi_lightcurve}d). In phase 1, flux of both polarities increased; however, increase of positive flux was more than that of negative flux. Flux of both polarities decayed at similar rate during phase 2. To have an understanding of flaring activity in the AR, we have plotted the variation of GOES SXR flux along with AIA 94 \AA\ light curve from 20 June 2015 12:00 UT to 21 June 2015 05:00 UT in Figure \ref{hmi_lightcurve}e. We find that, during the selected interval, there was no appreciable enhancement of SXR and EUV fluxes prior to the onset of the M-class flare on 21 June 2015 at $\approx$01:05 UT.

\subsection{Sunspot Rotation and Moving Magnetic Features}
\begin{figure}
\includegraphics[width=1\textwidth]{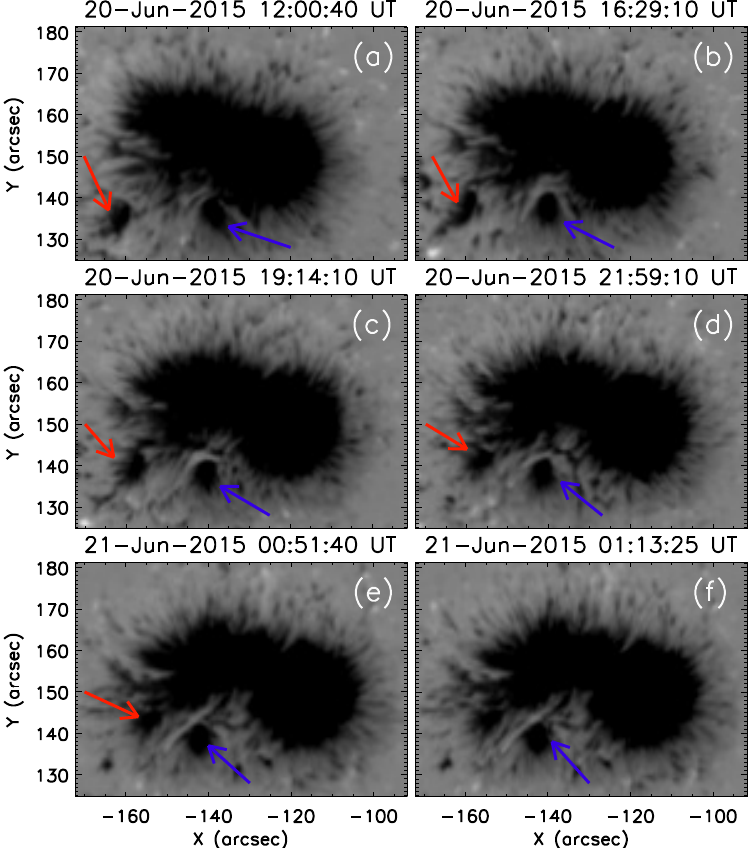}
\caption{Series of LOS magnetograms showing the evolution of the leading sunspot group of the AR (shown within the red box in Figure \ref{ar_phot}b). We have identified a small, nearly circular patch that exhibited significant clockwise rotation (indicated by the blue arrows) and another small magnetic patch displayed converging motion towards the sunspot (indicated by the red arrows).}
\label{mag_rot}
\end{figure}

The AR NOAA 12371 underwent significant morphological changes prior to the reported flare which include MMFs, emerging/cancelling flux elements, and sunspot rotation. To highlight these features, we have identified two sub-regions in the AR with most prominent photospheric changes which are marked by the red and blue coloured boxes in Figure \ref{ar_phot}b. The leading sunspot group of the AR (inside the red box in Figure \ref{ar_phot}b) was associated with a very interesting display of morphological evolution (see the animation associated with Figure \ref{ar_phot}). We note that, the extension of the leading sunspot group along the east-west direction increased from $\approx$55$''$ at 20 June 2015 12:00 UT to $\approx$65$''$ prior to the onset of the flare. During the same time, the north-south extension of the leading sunspot group decreased from $\approx$40$''$ to $\approx$35$''$ (\textit{cf.} Figures \ref{mag_rot}a and f). Further, a small circular element of negative polarity (indicated by the blue arrows in Figure \ref{mag_rot}) exhibited continuous motion along the southern boundary of the leading sunspot group. The diagonal expansion of the region alongside the motions of the small patch suggest ``clockwise rotation'' of the overall leading sunspot group. We also identify a distinct patch of small magnetic region (indicated by the red arrows in Figure \ref{mag_rot}) which, after initially exhibiting constant converging motion toward the major sunspot, merged with it during the final hours of 20 June 2015 (Figure \ref{mag_rot}e).

\begin{figure}
\includegraphics[width=0.9\textwidth]{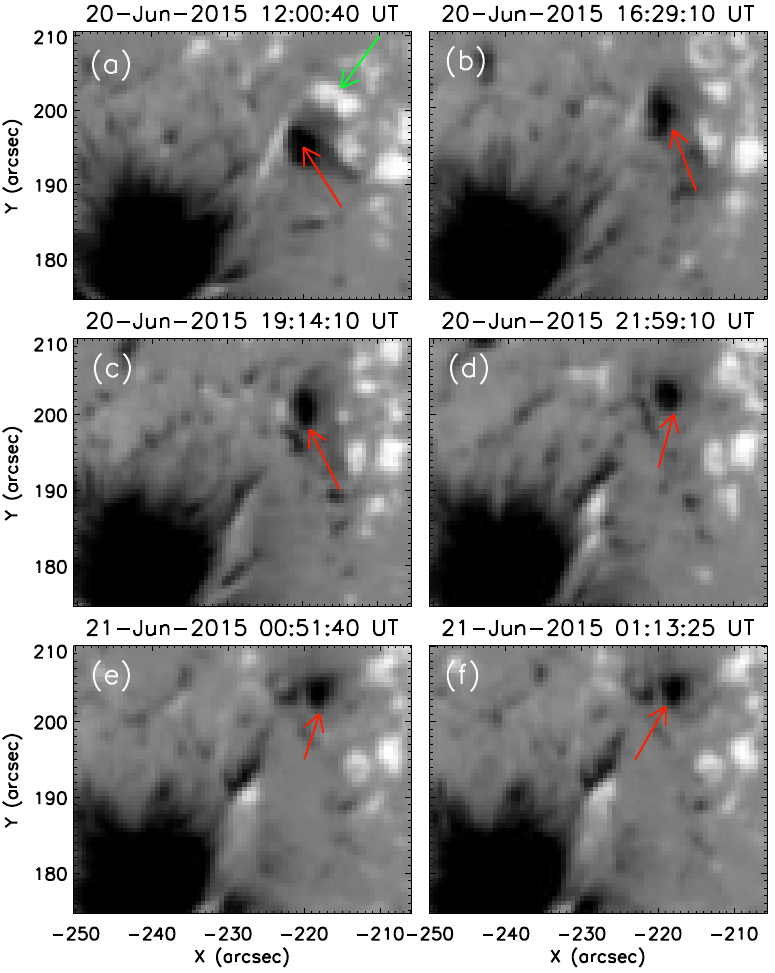}
\caption{Series of LOS magnetograms of the selected region shown within the blue box in Figure \ref{ar_phot}b showing striking and rapidly evolving MMFs along with flux emergence and cancellation. The red arrows indicate an MMF of a negative polarity. The green arrow indicates a region associated with cancellation of positive flux.}
\label{mag_mov}
\end{figure}

In Figure \ref{mag_mov}, we show the evolution of the region inside the blue box (Figure \ref{ar_phot}b) where we have indicated two particular features by red and green arrows. The red arrows indicate a striking MMF element of negative polarity which also shows significant morphological changes (see animation with Figure \ref{ar_phot}). The green arrows mark persistent cancellation of a positive flux element as the negative MMF moves toward north-west direction.

\section{Multi-wavelength Observations of the Eruptive Flare} \label{obs}
\subsection{Overview of the Event} \label{overview}
\begin{figure}
\includegraphics[width=1\textwidth]{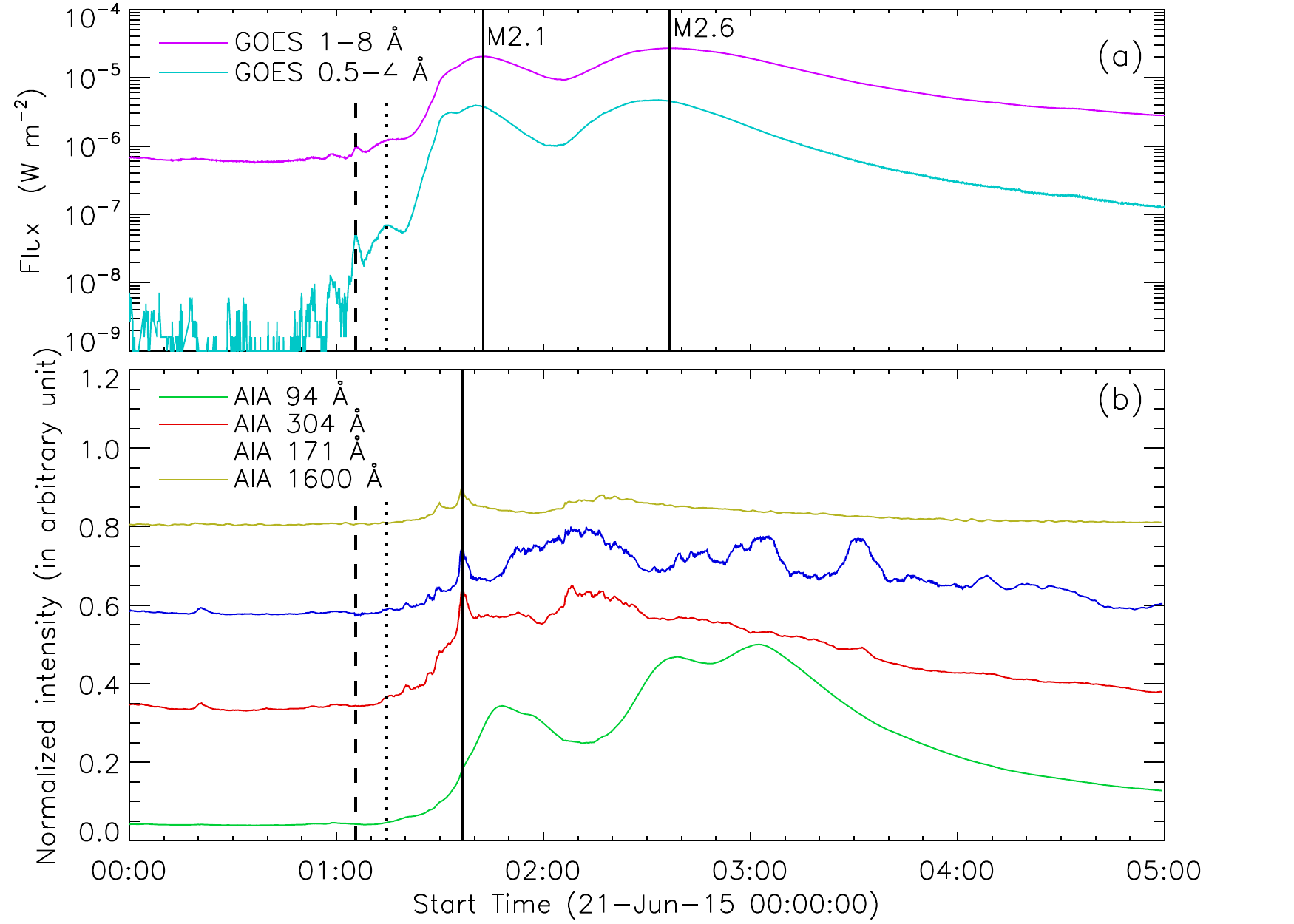}
\caption{(\textbf{a}): GOES SXR flux variation in the 1--8 \AA\ (magenta) and 0.5--4 \AA\ (green) channels showing the initiation and evolution of the dual-peak M-class flare on 21 June 2015. (\textbf{b}): Normalized intensity variations in the 94 \AA\ (green), 304 \AA\ (red), 171 \AA\ (blue), and 1600 \AA\ (yellow) channels of AIA during the flare. For better visualization, the AIA channels are scaled by factors of 0.5, 0.65, 0.8, and 0.9, respectively. The dashed and dotted lines in both the panels indicate two precursor events identified in the GOES SXR channels prior to the initiation of the main flare. The two peaks of the main flare are indicated by the solid lines in (\textbf{a}) during which GOES 1--8 \AA\ flux attained levels of M2.1 and M2.6, respectively. The solid line in (\textbf{b}) indicates an impulsive but short lived intensity enhancement observed in the AIA 304, 171, and 1600 \AA\ channels. An animation of this figure is provided in the supplementary materials.}
\label{lightcurve}
\end{figure}

The temporal evolution of the M-class flare on 21 June 2015 is depicted by the flux variation in the \textit{Geostationary Operational Environmental Satellite} (GOES) 1--8 \AA\ and 0.5--4 \AA\ channels which are plotted between 21 June 2015 00:00 UT and 05:00 UT in Figure \ref{lightcurve}a. The GOES profiles suggest that the eruptive flare evolved in two phases. According to the GOES 1--8 \AA\ time profile, the event started at $\approx$01:20 UT while the peaks of the two subsequent episodes of energy release were recorded at 01:42 and 02:36 UT during which the flux rose to the levels of M2.1 and M2.6, respectively. Further, the flare was associated with two brief pre-flare flux enhancements at $\approx$01:05 and $\approx$01:14 UT (indicated by the dashed and dotted lines, respectively, in Figure \ref{lightcurve}). Profile of GOES 0.5--4 \AA\ channel clearly shows that, while the flux enhancement during the first pre-flare peak was quite impulsive and short-lived, it was relatively gradual during the second pre-flare peak. Based on the temporal and spatial characteristics (discussed in the next subsection) of the episodic pre-flare enhancements, we refer them as SXR precursors.

In Figure \ref{lightcurve}b, we display the intensity variation of AIA (E)UV channels on 21 June 2015 from 00:00 UT to 05:00 UT. We note that, none of the AIA intensity profiles exhibited enhancement during the first GOES precursor while a subtle enhancement was observed during the second precursor in the AIA 304 and 171 \AA\ channels (\textit{cf.} the dashed and dotted lines in Figure \ref{lightcurve}a and b). Intensity in all the AIA channels, except the 94 \AA\ channel, displayed a sharp peak at $\approx$01:36 UT (indicated by the solid line in Figure \ref{lightcurve}b) while the second peak in these AIA channels was rather gradual. The intensity variation in the AIA 94 \AA\ channel diverged from other AIA channels in the timing and extended duration of the peaks. Interestingly, the AIA 94 \AA\ channel displayed three distinct peaks, first and second ones of which were consistent with the GOES M2.1 and the GOES M2.6 peaks, respectively. The highest of the three AIA 94 \AA\ peaks occurred during the gradual phase of the M-class flare. The flare moved into the gradual phase after $\approx$03:00 UT in all the AIA (E)UV and GOES SXR channels.

\subsection{Two-phase Flare Emission and Flux Rope Eruption} \label{flare}
\begin{figure}
\includegraphics[width=1\textwidth]{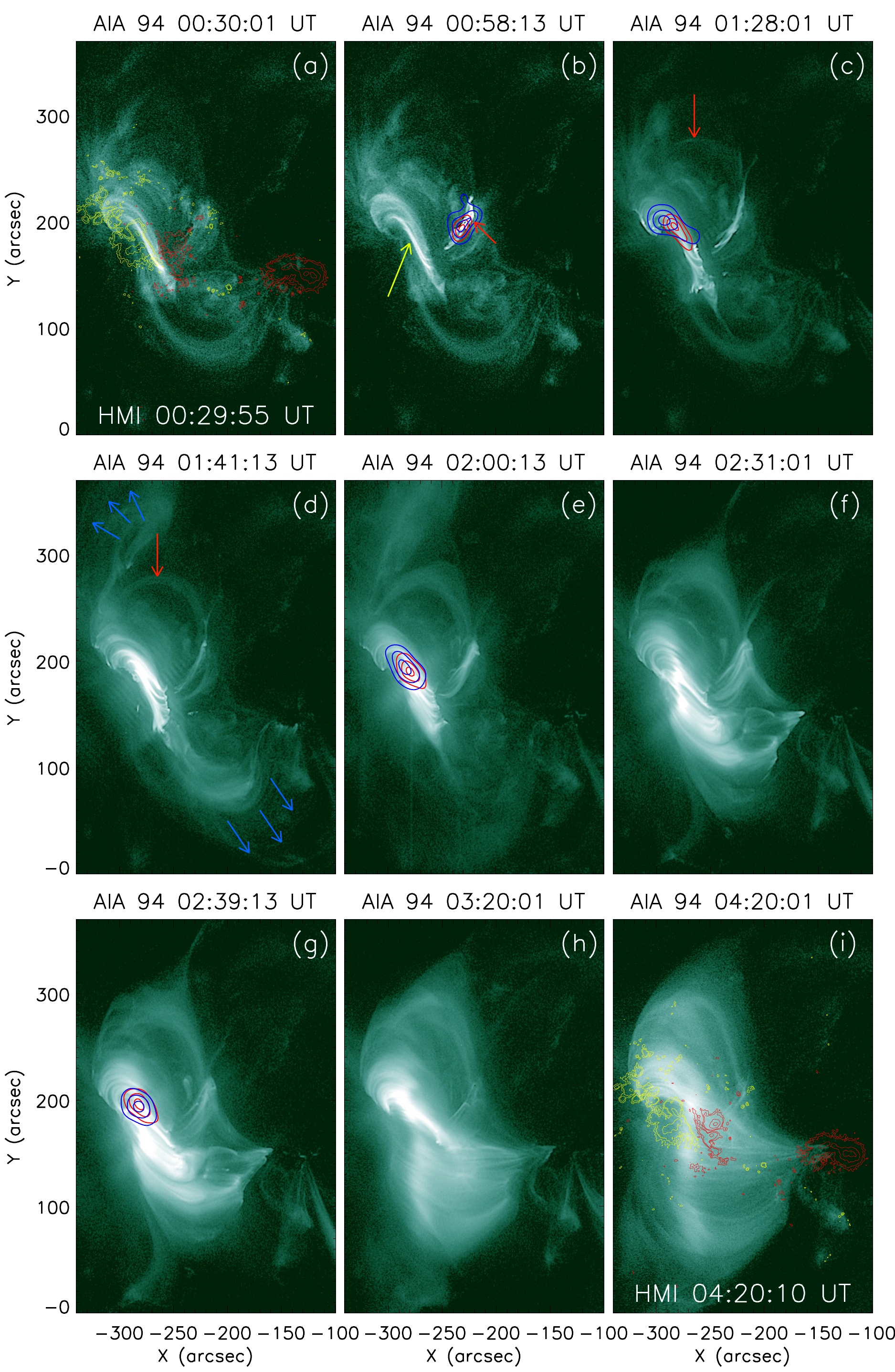}
\caption{Series of AIA 94 \AA\ images showing the evolution of the M-class flare on 21 June 2015 from the AR NOAA 12371. A distinct hot channel was observed during the pre-flare phase over the polarity inversion line in the AR which is marked by the yellow arrow in (\textbf{b}). A remote brightening observed prior to the onset of the flare is indicated by the red arrow in (\textbf{b}). The red arrows in (\textbf{c}) and (\textbf{d}) indicate a moving flash connecting the remote region and the northern leg of the hot channel. The blue arrows in (\textbf{d}) indicate the direction of erupting plasma during the impulsive phase of the flare. Co-temporal HMI LOS magnetogram contours are overplotted in (\textbf{a}) and (\textbf{i}) at $\pm(500,800,1500,2000)$ G. Red and yellow contours refer to negative and positive polarities, respectively. Co-temporal RHESSI contours in the energy bands 6--12 keV (red) and 12--25 keV (blue) are overplotted in selective panels. Contour levels are 60\%, 80\% and 95\% of the corresponding peak flux. All the images are derotated to 21 Jun 2015 00:30 UT.}
\label{fl94}
\end{figure}

In Figure \ref{fl94}, we plot a series of AIA 94 \AA\ images displaying the AR NOAA 12371 during different phases of the M-class flare. We readily observe the presence of a prominent hot channel at the core of the AR during the pre-flare phase (indicated by the yellow arrow in Figure \ref{fl94}b). Comparison of the location of the hot channel with the HMI LOS magnetogram contours in Figure \ref{fl94}a confirms that the hot channel was lying over the PIL in the trailing sunspot of the AR. After $\approx$00:52 UT, we observed a localized yet prominent brightening from a location near to the hot channel (indicated by the red arrow in Figure \ref{fl94}b). We note that, hard X-ray (HXR) emission of energies up to $\approx$25 keV originated from this location of pre-flare EUV brightenings. The examination of series of AIA 94 \AA\ images suggest that the brightness of this localized region initially increased up to $\approx$01:05 UT and then decreased till $\approx$01:10 UT before increasing again. These pre-flare episodic brightenings observed in AIA 94 \AA\ images are exactly co-temporal with the GOES SXR precursors observed at $\approx$01:05 UT and $\approx$01:14 UT (\textit{cf.} Figure \ref{lightcurve}a). A very interesting phenomena was observed around $\approx$01:28 UT in terms of anti-clockwise motion of brightness depicting a narrow semicircular path from the northern end of the region of precursor brightening to the northern leg of the hot channel. This moving flash in indicated by the red arrows in Figures \ref{fl94}c and d. The flare entered into the impulsive rise phase by $\approx$01:20 UT as the hot channel got activated. During this time, we noted HXR emission of energies up to $\approx$25 keV predominantly from the northern part of the hot channel. The progression of brightness from the adjacent precursor region to the northern leg of the hot channel was immediately followed by eruption of the hot channel, \textit{i.e.} the eruption was triggered. In Figure \ref{fl94}d, we indicate the direction of the hot channel eruption by the blue arrows. The eruption phase was followed by formation of post flare arcade in the trailing sunspot (Figures \ref{fl94}e and f) which are associated with HXR emission of energies up to $\approx$25 keV. A second phase of the eruption was observed between $\approx$01:53 UT and $\approx$02:05 UT followed by further restructuring of the AR loops at even larger scales, as inferred from the formation of large post flare arcade connecting the trailing sunspot with the leading sunspot. The large post-flare arcade was associated with strong diffused emission till $\approx$02:40 UT (Figure \ref{fl94}g) when the flare reached its second peak. The active region did not show any significant morphological change afterwards till the end of our studied period except the brightness of the large post flare arcade slowly reduced as the flare had moved into the gradual phase (Figures \ref{fl94}h, i). The above observations led \citet{Joshi2018} to conclude that two distinct phases of magnetic reconnection occurred successively at two separate locations and heights of the AR corona in the wake of single, large hot channel eruption. Based on the temporal and spatial proximity of the two distinct energy release phases along with spectral characteristics of emission, \citet{Lee2018} has termed the two-phases of energy release as the signature of a ``composite flare'' triggered by the flux rope eruption.

\begin{figure}
\includegraphics[width=1\textwidth]{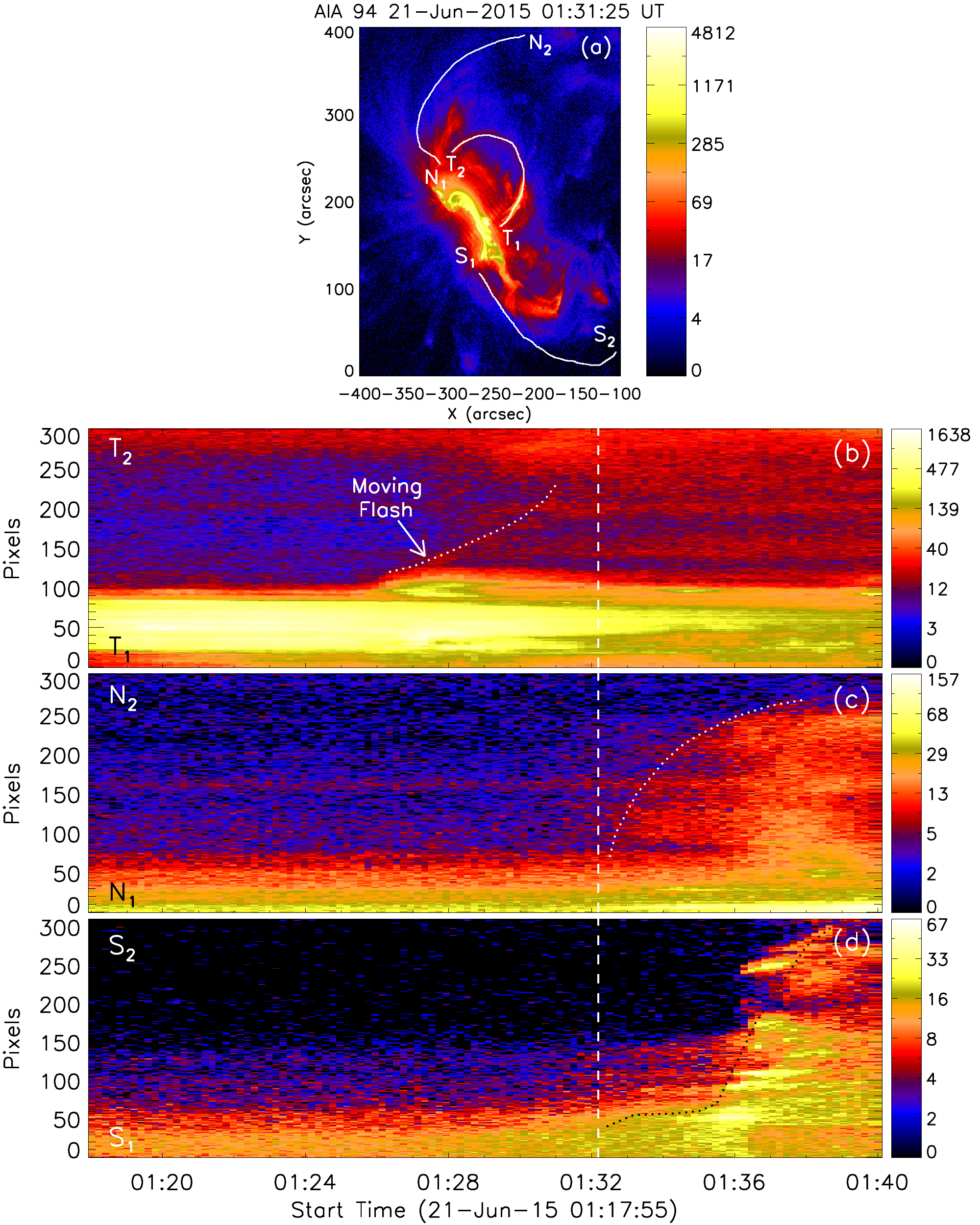}
\caption{(\textbf{a}): AIA 94 \AA\ images prior to the onset of the hot channel eruption. The three curves (marked by T$_1$T$_2$, N$_1$N$_2$, and S$_1$S$_2$) indicate three slits along which time-slice diagrams are computed. (\textbf{b})--(\textbf{d}): Time-slice diagrams corresponding to the slits T$_1$T$_2$, N$_1$N$_2$, and S$_1$S$_2$, respectively. The dotted curve in (\textbf{b}) indicate the anti-clockwise motion of the subtle brightness from the precursor location to the northern leg of the hot channel. Dotted curves in (\textbf{c}) and (\textbf{d}) highlight the eruptive motion of the hot channel. The dashed vertical line in (\textbf{b})--(\textbf{d}) indicate the estimated time of the arrival of the brightness to the northern leg of the hot channel along T$_1$T$_2$. An animation of this figure is provided in the supplementary materials.}
\label{triggering}
\end{figure}

To have a further clarification on the triggering of the hot channel eruption, we selected three slits and computed time-slice diagrams along them (Figure \ref{triggering}). These time-slice diagrams can be effectively used to observe the time evolution of plasma eruption and brightness progression along the selected slits (see also the animation associated with Figure \ref{triggering}). From Figure \ref{triggering}b, we find that the motion of brightness (apparent signatures of slipping reconnection) from the precursor location started at $\approx$01:26 UT. We have highlighted the motion of the brightness in Figure \ref{triggering}b by a dotted curve. The estimated time of the arrival of the brightness at the T$_2$ point is indicated by the dashed vertical line in Figures \ref{triggering}b--c. From Figures \ref{triggering}c and d, it becomes evident that the eruption started immediately after the progression of brightness reached the core of the AR, \textit{i.e.} the eruption was triggered by the processes linked with the moving brightness.

\begin{figure}
\includegraphics[width=1\textwidth]{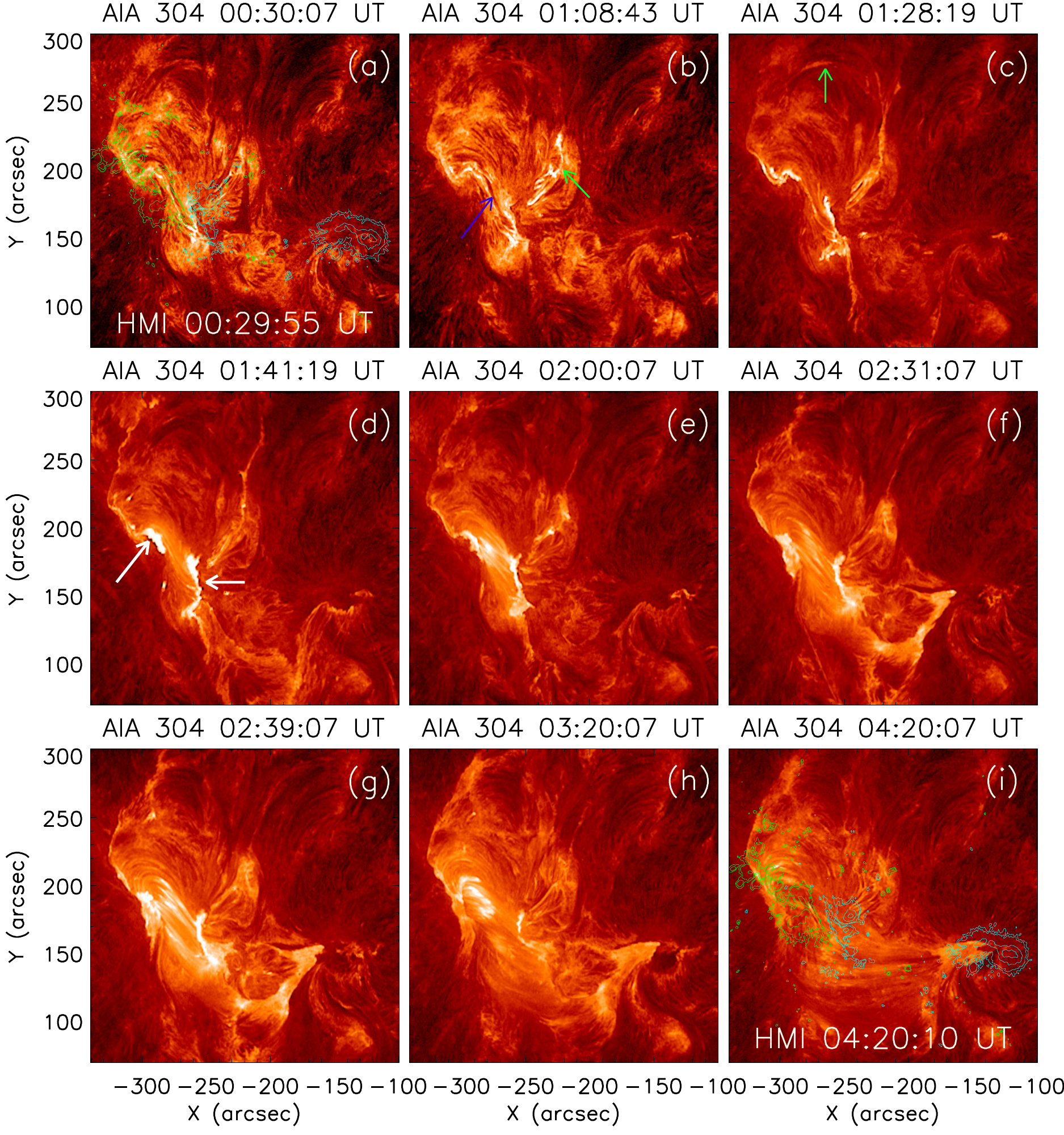}
\caption{Series of AIA 304 \AA\ images displaying different phases of the M-class flare. A filament was observed (indicated by the blue arrow in (\textbf{b})) which was co-spatial to the hot channel (\textit{cf.} Figure \ref{fl94}b). The remote brightening and the moving flash are indicated by the green arrow in (\textbf{b}) and (\textbf{c}), respectively. The white arrows in (\textbf{d}) mark the two flare ribbons during the impulsive phase of the flare. Co-temporal HMI LOS magnetogram contours are overplotted in (\textbf{a}) and (\textbf{i}) at $\pm(500,800,1500,2000)$ G. Green and sky contours refer to positive and negative polarity, respectively. All the images are derotated to 21 Jun 2015 00:30 UT.}
\label{fl304}
\end{figure}

In Figure \ref{fl304}, we display a series of AIA 304 \AA\ channel showing the evolution of the AR during the M-class flare. A filament (marked by the blue arrow in Figure \ref{fl304}b) was observed to lie along the PIL in the trailing sunspot region (\textit{cf.} Figure \ref{fl304}b with HMI LOS contours in Figure \ref{fl304}a) which is co-spatial with the location of the hot channel observed in the AIA 94 \AA\ channel images (\textit{cf.} Figures \ref{fl304}b and \ref{fl94}b). The adjacent precursor activity was observed in the AIA 304 \AA\ images also (Figure \ref{fl304}b and c). During the impulsive phase of the M-class flare, a clear set of flare ribbons formed in the trailing sunspot region (indicated by the arrows in Figure \ref{fl304}d). As expected from the standard flare scenario, the separation between the two ribbons increased, albeit rather slowly, with time and a dense post flare arcade was eventually formed connecting the two ribbons (Figures \ref{fl304}d--h). 

\subsection{Small-scale Pre-eruption Processes} \label{preflare}
\begin{figure}
\includegraphics[width=1\textwidth]{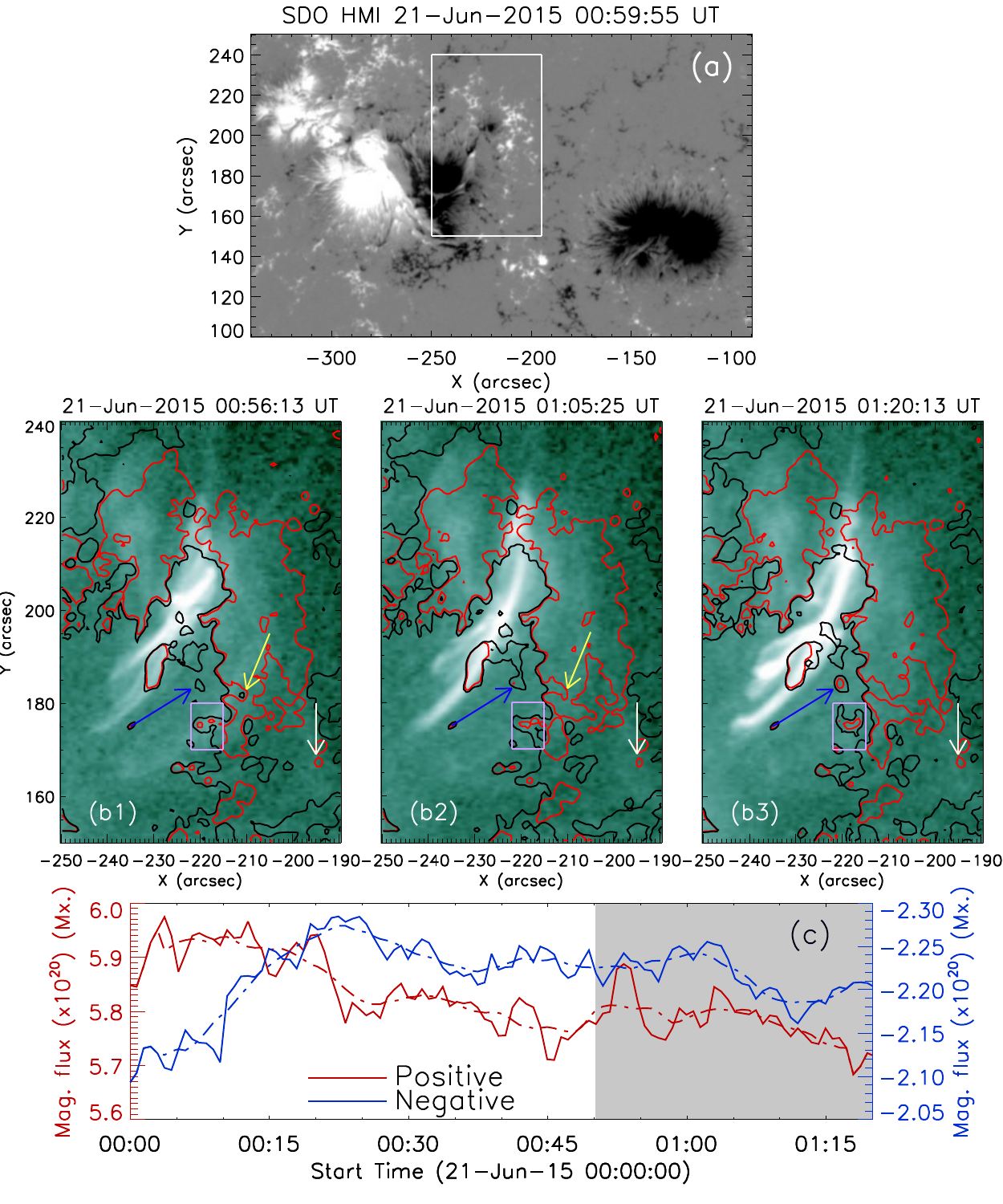}
\caption{(\textbf{a}): HMI LOS magnetogram of AR 12371 prior to the onset of the M-class flare reported in this article. The box in (\textbf{a}) outlines the region that displayed small-scale magnetic field changes along with MMFs. (\textbf{b1})--(\textbf{b3}): AIA 94 \AA\ images of the region, shown within the box in (\textbf{a}), are overplotted with co-temporal LOS magnetograms. Black and red contours refer to negative and positive polarities, respectively. Contour levels are $\pm25$ G. The arrows and the boxes in these panels indicate different instances of flux emergence and decay in this region prior to the triggering of the flare. All the images in (\textbf{a}) and (\textbf{b1})--(\textbf{b3}) are derotated to 21 Jun 2015 01:00 UT. In (\textbf{c}), we display the variation of magnetic flux in the region within the box in (\textbf{a}) on 21 Jun 2015 from 00:00 UT up to the onset of the flare at $\approx$01:20 UT. The shaded region in (\textbf{c}) denotes the interval of precursor enhancements observed in GOES SXR channels.}
\label{act}
\end{figure}

From AIA EUV images (Figures \ref{fl94} and \ref{fl304}) it is clearly understood that the earliest flare brightening occurred at the west of the pre-existing hot channel, (\textit{i.e.} MFR) which evolved with time but remained within a localized region. As discussed earlier, we identify this brightening as GOES SXR precursor (see Figure \ref{lightcurve}a). This was followed by the activation and eruption of the hot channel as a sequence of activities. We recall that this region region of precursor brightening was associated with rapidly evolving dispersed magnetic field of positive and negative polarities in which MMFs were observed besides emergence and cancellation of magnetic flux (Figure \ref{mag_mov}). The precursor activities and its relation to the small-scale magnetic field changes are further analysed in Figure \ref{act}. In Figure \ref{act}a, we highlight the region of precursor brightening over the HMI magnetogram by the box and show co-temporal overplots of the AIA 94 \AA\ images with the magnetograms in Figures \ref{act}b1--b3. We identify several instances of flux emergence and cancellation of both polarities  which are indicated by the arrows of different colors and the boxes in Figures \ref{act}b1--b3. In Figure \ref{act}c, we show the evolution of LOS magnetic flux from the region shown within the box in Figure \ref{act}a on 21 June 2015 from 00:00 UT till the onset of the impulsive phase of the flare at $\approx$01:20 UT. We find that magnetic flux of both positive and negative polarities exhibited episodic increase and decrease from the region, further implying significant small-scale flux variations. Notably, the overall trend of positive flux in this region displayed a slow decay while the negative flux increased initially up to $\approx$00:25 UT and decayed gradually thereafter.

\subsection{Morphology and Evolution of Photospheric Longitudinal Current} \label{ph_curr_fe}
\begin{figure}
\includegraphics[width=1\textwidth]{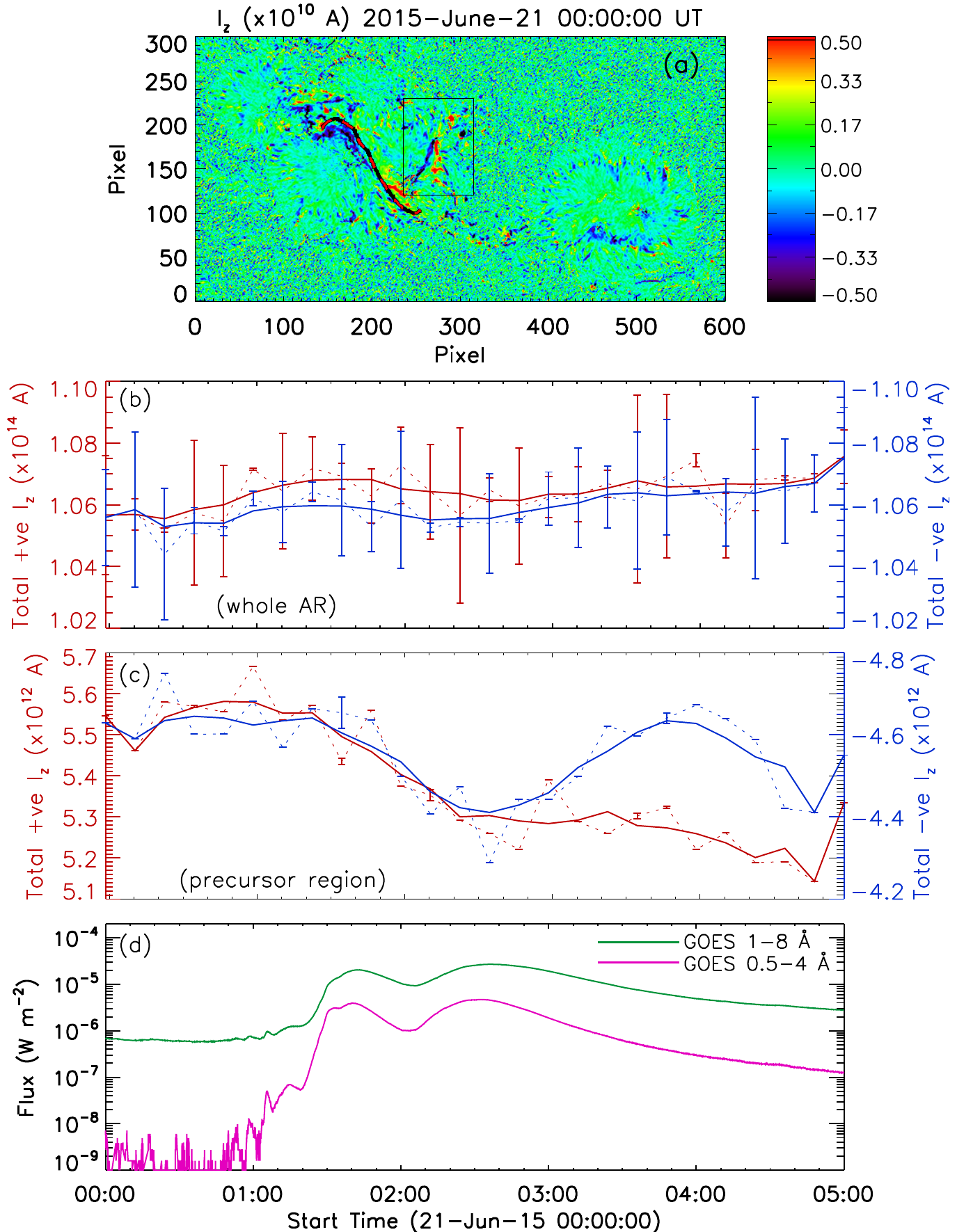}
\caption{(\textbf{a}): Distribution of longitudinal component of the photospheric current in AR 12371 on 21 June 2015 00:00 UT. The region associated with precursor brightening is enclosed by the box in (\textbf{a}). Note that, for better visualization, we have saturated the $I_\textrm{z}$ values at $\pm$0.5$\times$10$^{10}$ A in (\textbf{a}). The approximate location of the hot channel is indicated by the red-black dashed line in (\textbf{a}). In (\textbf{b}) and (\textbf{c}), we plot the variation of vertical component of photospheric current within the entire AR and within the box in (\textbf{a}), respectively. The vertical bars represent 1$\sigma$ uncertainty in the calculation. For comparison, we have plotted the variation of GOES SXR channels in (\textbf{d}).}
\label{j_var}
\end{figure}

Electric current density on the photosphere, being a direct consequence of flux emergence and decay as well as photospheric motions, is expected to provide important insights toward understanding the onset of flares. The longitudinal component of current density ($j_\textrm{z}$) on the photosphere can be calculated from horizontal components of magnetic field ($B_\textrm{x}$ and $B_\textrm{y}$) using the Ampere's law \citep{Tan2006, Kontogiannis2017}:
\begin{equation}
j_\textrm{z}=\frac{1}{\mu_\circ}(\frac{\textrm{d}B_\textrm{y}}{\textrm{d}x}-\frac{\textrm{d}B_\textrm{x}}{\textrm{d}y})
\end{equation}
From current density ($j_\textrm{z}$), we derive current ($I_\textrm{z}$) by multiplying $j_\textrm{z}$ with the area of one pixel, \textit{i.e.} $\approx$13.14$\times$10$^{10}$ m$^2$. In Figure \ref{j_var}, we plot the temporal evolution of average $I_\textrm{z}$ for the overall active region (Figure \ref{j_var}b) as well as the location of precursor brightening (Figure \ref{j_var}c). From the spatial distribution of $I_\textrm{z}$ prior to the hot channel activation (Figure \ref{j_var}a), we find large concentration of positive and negative currents along the narrow strip delineated by the PIL with the maximum and minimum values of $I_\textrm{z}$ in the active region being 2.09$\times$10$^{10}$ A and -2.24$\times$10$^{10}$ A, respectively. It is noteworthy that the region displaying precursor brightenings in the corona and underlying MMFs in the photosphere exhibited a complex distribution of $I_\textrm{z}$. We have indicated this region in the box in Figure \ref{j_var}a and identify this as the triggering region. We find that in the overall AR, both positive and negative currents experienced a slow variation. Both the positive and negative components of $I_\textrm{z}$ slowly increased during the pre-flare phase of the flare and decreased once the flare onset took place.

\begin{figure}
\includegraphics[width=1\textwidth]{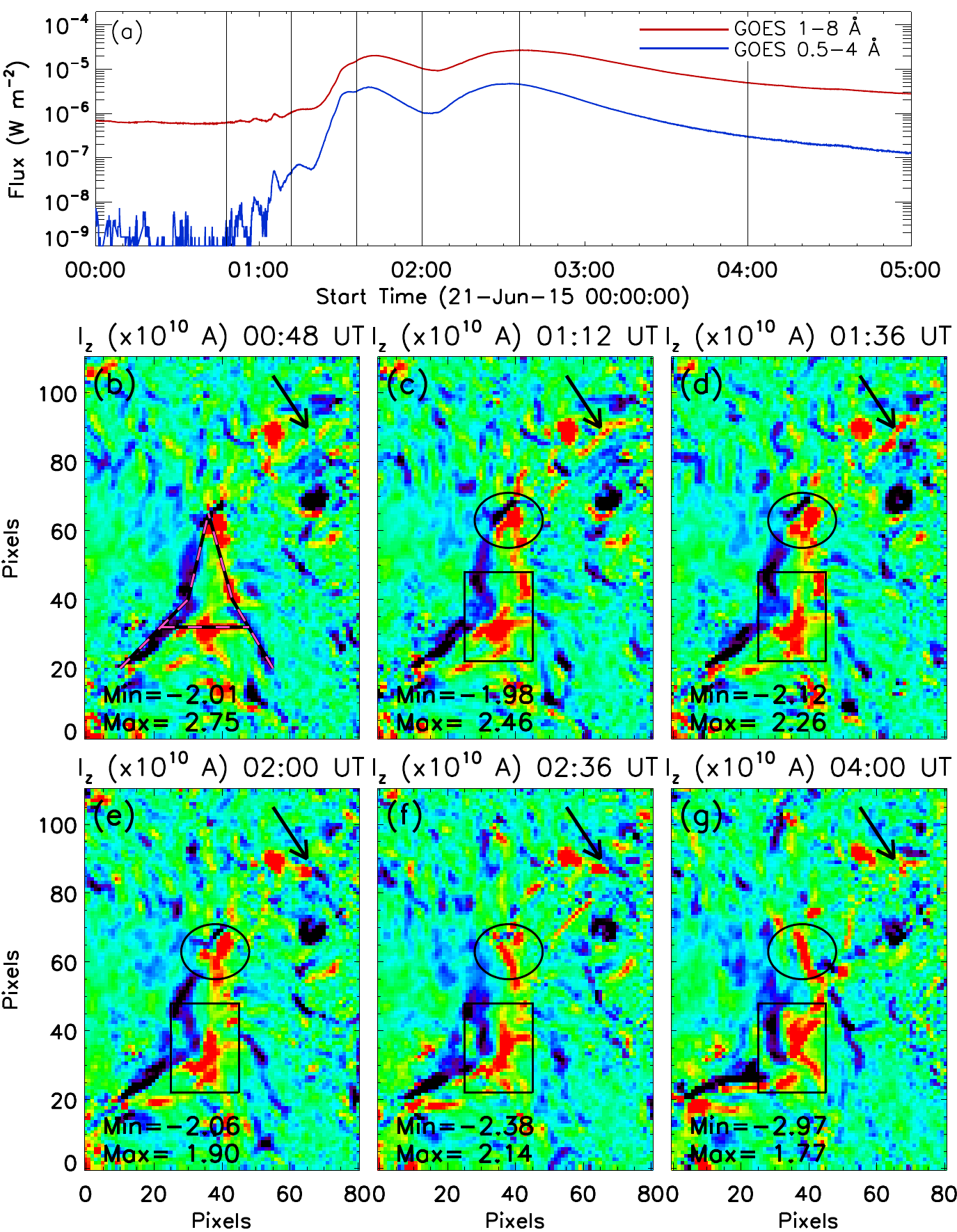}
\caption{(\textbf{a}): GOES SXR light curves showing the evolution of the M-class flare. In (\textbf{b})--(\textbf{g}), we plot the distribution of vertical component of photospheric current over the region shown within the box in Figure \ref{j_var}a at six different times as indicated by the black lines in (\textbf{a}). Notably, on a whole, few regions of strong current of opposite polarity constitute a structure similar to the English letter ``A'' which is outlined by the black-pink dashed lines in (\textbf{b}). We highlight few major changes in the distribution of current by the arrow, oval and box in (\textbf{c})--(\textbf{g}). For better visualization, values of $I_\textrm{z}$ are saturated at $\pm$0.5$\times$10$^{10}$ A. Maximum and minimum values of $I_\textrm{z}$ with order of 10$^{10}$ A within the selected FOV are indicated in each of these panels.}
\label{j_img}
\end{figure}

In Figure \ref{j_img}, we show the evolution of the spatial distribution of $I_\textrm{z}$ in the triggering region of the AR (within the box in Figure \ref{j_var}a). For convenience, we have plotted GOES SXR light curves in Figure \ref{j_img}a where the timings of the Figures \ref{j_img}b--g are indicated by the vertical lines. We find that, in the triggering location, small-scale regions of both positive and negative $I_\textrm{z}$ were mostly distributed randomly. However, an interesting structure of the shape ``A'' formed by $I_\textrm{z}$ was very clear in the region during the pre-flare phase (outlined by pink-black dashed lines in Figure \ref{j_img}b). In the ``A'' shaped distribution, the left arm was completely made of negative $I_\textrm{z}$ while the right arm was consisted of the positive $I_\textrm{z}$ in the northern part and negative $I_\textrm{z}$ in the southern part. The connecting part of the arms in the ``A'' was consisted of positive $I_\textrm{z}$. During the SXR precursor enhancement, in the northern tip of the structure (outlined by the oval shape in Figures \ref{j_img}b--g), $I_\textrm{z}$ of opposite polarities became very close to each other (Figures \ref{j_img}b and c) which may be ideal for dissipation of current in the form of magnetic reconnection. As Figures \ref{j_img}d--g suggest, with the evolution of the flare, strength of $I_\textrm{z}$ at the tip of the ``A'' significantly decreased and the left arm of the ``A'' became fragmented (see region inside the box in Figures \ref{j_img}b--g). We spotted another interesting feature from Figures \ref{j_img}b--g in the form of appearance and decay of a positive current region which we indicate by the black arrows.

\section{Non-linear Force Free Field Extrapolation Results} \label{sec_nlfff}
\subsection{Modelled Coronal Magnetic Configuration} \label{sec_extrp}
\begin{figure}
\includegraphics[width=1\textwidth]{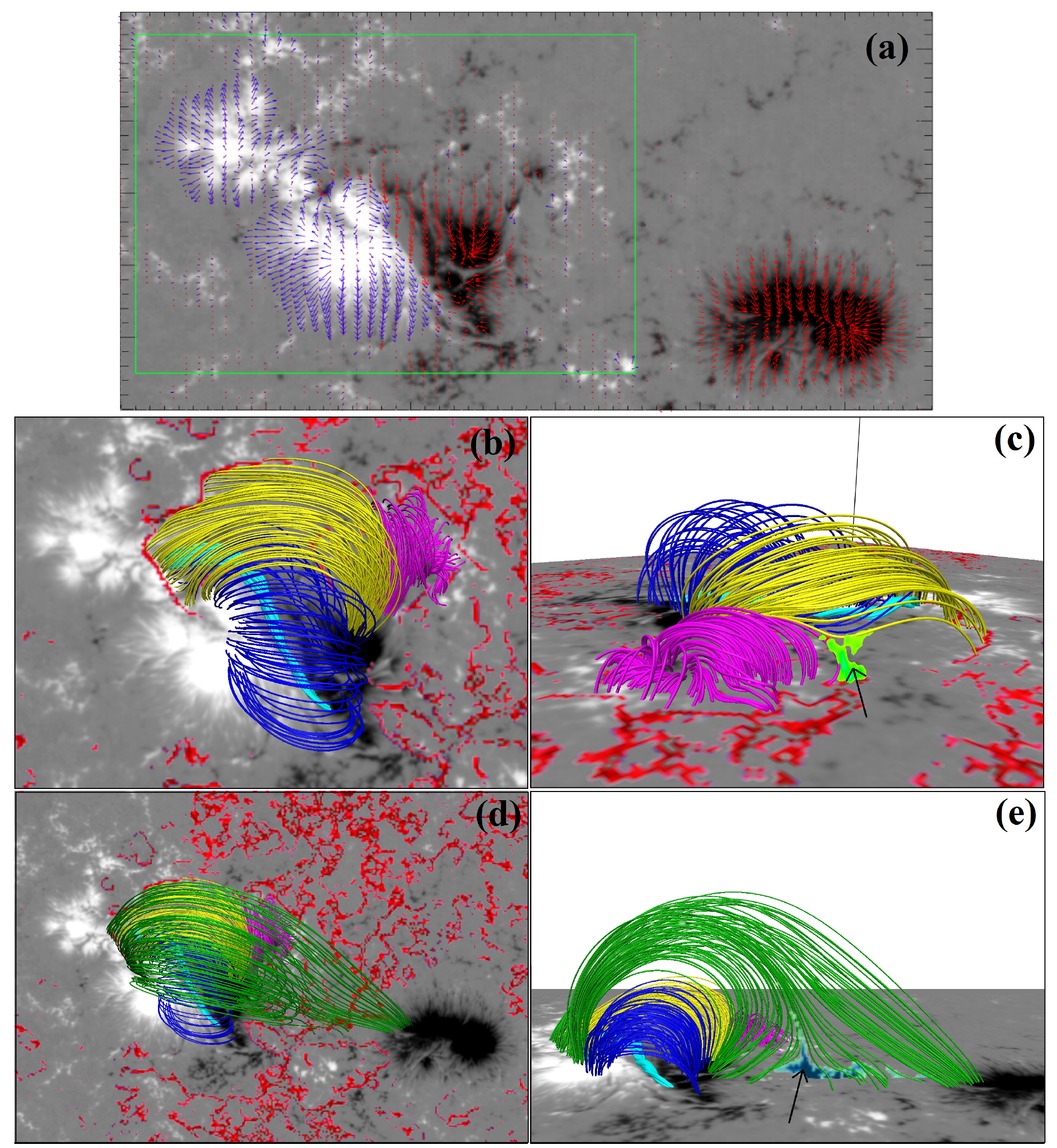}
\caption{(\textbf{a}): CEA projected vertical component ($B_\textrm{z}$) of the HMI vector magnetogram of the AR NOAA 12371 at 01:00 UT on 21 June 2015. Blue and red arrows over the magnetogram indicate the horizontal component of magnetic field associated with positive and negative ($B_\textrm{z}$), respectively. (\textbf{b})--(\textbf{e}): Modelled coronal magnetic configuration based on the NLFFF extrapolation results of the vector magnetogram shown in (\textbf{a}). Different sets of NLFFF field lines associated with the trailing sunspot group, \textit{i.e.} within the green box in (\textbf{a}), from top and side views, are shown in (\textbf{b}) and (\textbf{c}), respectively. The neon-green coloured region (also marked by the black arrow) indicate regions characterised by \textit{log(Q)}$>$9. In (\textbf{d}) and (\textbf{e}), we show the top and side view of the model lines in the whole FOV as in (\textbf{a}). The blue regions (also marked by the black arrow) in (\textbf{d}) indicate a region possessing high \textit{Q} value (\textit{log(Q)}$>$8). The background white boundary behind the green lines in (\textbf{d}) direct to the north. In all the panels, photospheric regions shown by red color are characterised by \textit{log(Q)}$>$7.}
\label{nlfff}
\end{figure}

In order to understand the coronal connectivities between the photospheric magnetic polarities of the AR, we performed a non-linear force free field (NLFFF) extrapolation. For the purpose, we selected an HMI magnetogram at 01:00 UT on 21 June 2015 which represents the photospheric configuration prior to the onset of eruption (Figure \ref{nlfff}a). In Figures \ref{nlfff}b and c, we display different sets of NLFFF lines associated with the trailing sunspot group, from top and side views, respectively. The NLFFF extrapolation results readily suggest the presence of an extended flux rope (shown by sky-coloured lines) over the PIL which was enveloped by a set of low coronal closed loops (shown by blue lines) connecting the opposite polarity regions of the trailing sunspot group. Importantly, the modelled flux rope is situated at the same location where the hot coronal channel was identified in the AIA images (\textit{cf.} Figure \ref{nlfff}b and \ref{fl94}a). The coronal configuration associated with the precursor brightening region is displayed by the pink lines in Figure \ref{nlfff}. Further, we identified a set of closed field lines (shown by the yellow lines) connecting the opposite polarity regions in the trailing sunspot group, a part of which are situated at a very close proximity to the pink lines (see Figure \ref{nlfff}c). Notably, the anti-clockwise motion of the brightness from the precursor location, clearly revealed by the time-slice diagram shown in Figure \ref{triggering}a, matches reasonably well with the footpoints of the yellow lines. We also note that, a few of the yellow and pink lines displayed a drastic change in the field-line linkage. In literature, such structures are termed as quasi-separatrix layers (QSLs) and are characterised by high squashing factor \citep[\textit{Q}; see,][]{Priest1995}. The \textit{Q} value of this region is found to be higher than 10$^9$ (shown by the neon-green coloured patch and the black arrow in Figure \ref{nlfff}c).

To have a further insight of the anti-clockwise motion of the brightening prior to the eruption of the hot channel, we show the photospheric regions with a \textit{Q} value greater than 10$^7$ by red colour in Figures \ref{nlfff}b--e. We find that, the footpoints of the yellow lines perfectly match with regions of high \textit{Q} values (Figure \ref{nlfff}c) which implies that the anti-clockwise motion of brightness from the precursor location was a slipping reconnection \citep[see \textit{e.g.},][]{Demoulin1997, Craig2014, Janvier2016}.

In Figures \ref{nlfff}d and e, we display the coronal connectvities in the whole AR from top and side views, respectively. We note the presence of large coronal loops that connected the positive regions of the trailing sunspot group to the leading negative sunspot group (shown by green lines in Figure \ref{nlfff}d--e). Also, a part of the green lines originating at the positive polarity region were connected with the adjacent negative polarity region of the same sunspot group. These two sets of green lines constituted a second QSL (\textit{log(Q)}$>$8) which we have indicated by a black arrow and the blue coloured patch in Figure \ref{nlfff}d. The observations of spatial progression of brightness within the same region (Figures \ref{fl94}f--g, \ref{fl304}f--h and animation associated with Figure \ref{lightcurve}) are consistent with the scenario of slipping reconnection. Interestingly, the same region of high \textit{Q} value was associated with a large-scale slipping reconnection event during the M-class flare on 22 June 2015 \citep{Jing2017} which implies that the large-scale magnetic structure associated with the active region remained preserved on the next day despite the eruption of a halo CME on 21 June 2015.

\subsection{Evolution of Magnetic Free Energy} \label{sec_fe}
\begin{figure}
\includegraphics[width=1\textwidth]{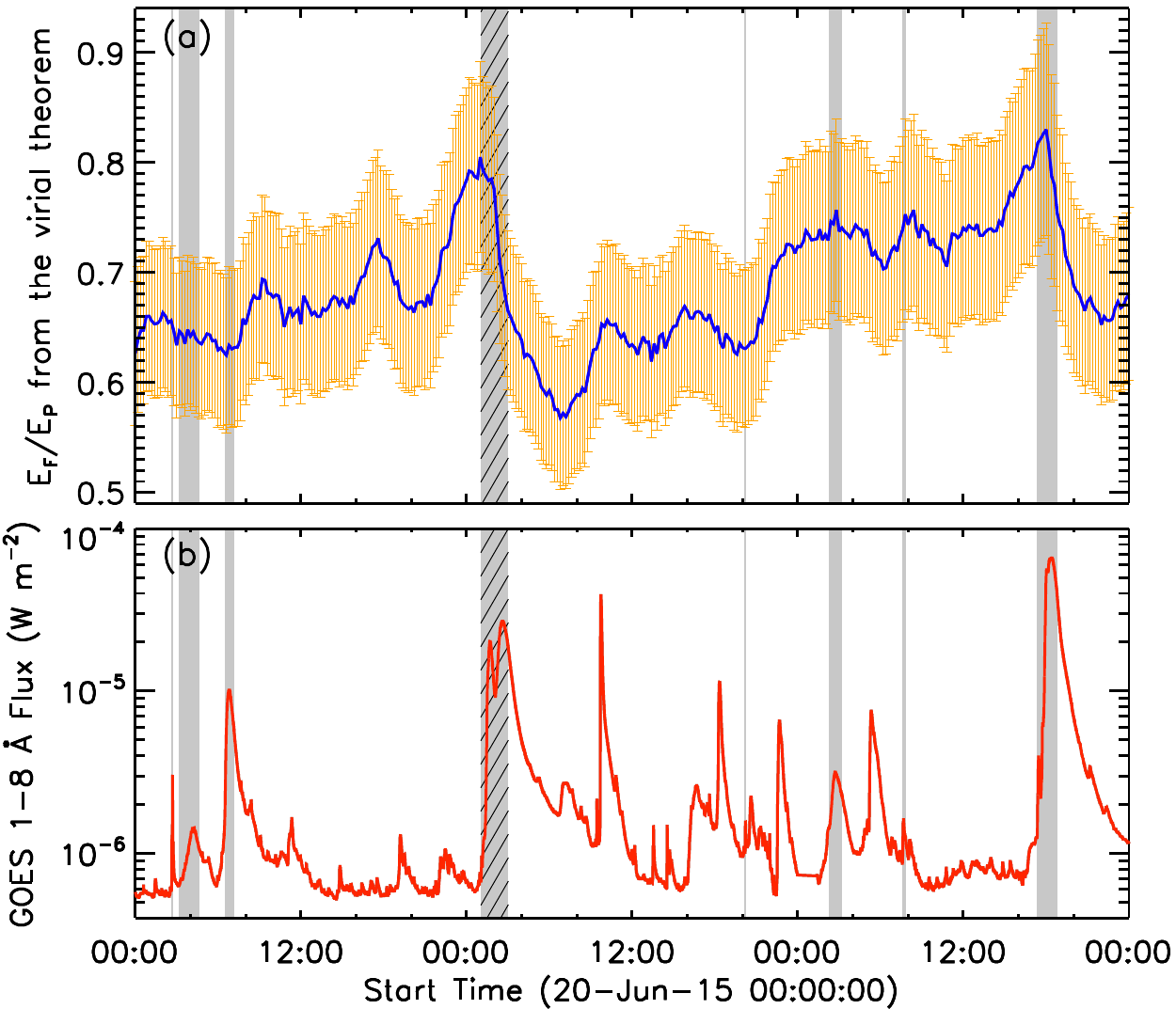}
\caption{(\textbf{a}): Evolution of magnetic free energy during 20--22 June, 2015, calculated by virial theorem. 1$\sigma$ error in the calculation of free energy is plotted by yellow bars. For comparison, GOES 1--8 \AA\ SXR flux variation is plotted in (\textbf{b}). The shaded intervals represent the durations of C-, and M-class flares originated from AR NOAA 12371. The striped dashed interval indicates the duration of the flare reported in this article.}
\label{imfe}
\end{figure}

The magnetic free energy ($E_\textrm{F}$) associated with an AR is important in order to understand the energy budget of the flares originated from that AR. $E_\textrm{F}$ can be estimated by the formula: 
\begin{equation}
E_\textrm{F}=E_\textrm{N}-E_\textrm{P}=\int_v\frac{{B_\textrm{N}}^2}{8\pi}\textrm{d}v-\int_v\frac{{B_\textrm{P}}^2}{8\pi}\textrm{d}v
\end{equation}
where $E_\textrm{N}$ and $E_\textrm{P}$ are non-potential energy and potential energy, respectively. We have calculated magnetic energy stored in the active region NOAA 12371 by employing the magnetic virial theorem \citep{Klimchuk1992}. According to this theorem, the magnetic energy stored in a coronal force-free magnetic field is given by the surface integral at the photospheric boundary involving the three vector magnetic field components, \textit{i.e.}
\begin{equation}
E=\frac{1}{4\pi}\int_{z=0}(xB_\textrm{x}+yB_\textrm{y})B_\textrm{z}\textrm{d}x\textrm{d}y
\end{equation}
where $B_\textrm{x}$, $B_\textrm{y}$, and $B_\textrm{z}$ are the $x$-, $y$-, and $z$-components of the photospheric magnetic field, respectively. We obtained the 3 components of photospheric magnetic field from the vector magnetograms of the ``hmi.sharp\_cea\_720s'' series. The magnetograms were then pre-processed as described in \citet{Wiegelmann2006}. We plot the evolution of the free magnetic energy stored in the AR NOAA 12371 normalised by the corresponding potential energy in Figure \ref{imfe}a during the most active phase of AR 12371 (20-22 June, 2015). For reference, we have plotted GOES 1--8 \AA\ SXR flux for the whole duration in Figure \ref{imfe}b. In this duration, the AR produced 3 M-class flares: class M1.0 on 20 June, class M2.6 on 21 June and class M6.5 on 22 June (see Table \ref{table1}). From Figure \ref{imfe}a, we find that prior to the M2.6 and M6.5 class flares, magnetic free energy in the AR 12371 was over 80\% of the corresponding potential energy. After both the flares, free energy decreased and reached local minima at $\approx$57\% and $\approx$65\% of the corresponding potential energies, respectively. Interestingly, we did not find any significant decrease in magnetic free energy during and after the M-class flare on 20 June, neither free energy increased prior to the flare.

\section{Discussion} \label{discus}
We present a multi-wavelength analysis of the flux rope eruption on 21 June 2015 from the active region NOAA 12371 with a particular emphasis on its triggering mechanism. The automated catalogue of GOES flares\footnote{see https://www.swpc.noaa.gov/products/solar-and-geophysical-activity-summary.} has enrolled two flares on 21 June 2015. However, from the evolution of the active region from the hot channel to the formation of large post-flare arcade during the two GOES peaks, we can convincingly call the GOES flaring events as flux rope eruption, \textit{i.e.} a dual-peaked long duration event.

The temporal evolution of the GOES SXR flux suggests that the flare was associated with two-stage precursor emission. Comparison of GOES SXR fluxes with AIA observations confirms that both the episodes of the subtle flux enhancements prior to the flare were caused by localised brightenings in a nearby region situated to the west of the hot channel (Figures \ref{fl94}, \ref{fl304}, and \ref{act}). These localised brightenings indicate energy release at small-scale within the same active region that can provide potential trigger for eruption by destabilizing a stable MFR \citep{Farnik1996, Farnik1998, Warren2001, Farnik2003, Sterling2005, Farnik2003, Chifor2007, Kim2008, Liu2009, Joshi2011, Joshi2016, Joshi2017, Woods2017, Dhara2017, Woods2018, Hernandez2019, Mitra2019}. Based on the location of small-scale energy release compared to main flaring event, pre-flare events can be categorized in three classes: co-spatial, adjacent/overlapping, and distant \citep{Farnik1996, Farnik1998}; the present case being an example of adjacent pre-flare activity. Further, the temporal and spatial evolution of the event suggests that pre-flare emission essentially acted as the precursor to the main eruption.

An important aspect of this study is to investigate the origin of the pre-flare activities and their spatial relation with the eruption of the MFR. NLFFF extrapolation results (Figure \ref{nlfff}) suggests that the triggering region, (\textit{i.e.} the region of precursor activity) was associated with coronal magnetic loops with high shear (pink lines in Figure \ref{nlfff}). \citet{Zhang1995} demonstrated that sheared magnetic field leads to the formation of electric current on the photosphere which can be accompanied by flares. \citet{Tan2006} investigated evolution of photospheric current during two flares of classes M1.0 and M8.7 occurring from two different ARs. Although, both the ARs were subjected to rapid flux emergence, the two flares differed significantly in the evolution of photospheric longitudinal current. Their analysis revealed that for the M1.0 flare, the longitudinal electric current density dropped rapidly; while it increased for the case of the M8.7 flare. They concluded rapid emergence of current carrying flux to be responsible for the increasing longitudinal current during the M8.7 flare while their explanation for the decrease of electric current for the M1.0 flare the was dissipation of magnetic free energy in the solar atmosphere. Our analysis suggests that in the triggering location, a few localized regions with high values of vertical component of photospheric electric current with opposite polarities were situated very close to each other (Figure \ref{j_img}). During the GOES pre-flare peaks, one particular set of regions with opposite polarity longitudinal current became adjacent to each other (within the oval in Figures \ref{j_img}b, c) suggesting that the initial reconnection most likely occurred from this location. Once the reconnection began, it induced further reconnection events in the nearby stressed magnetic field lines of the region resulting in the enhancement of plasma temperature. These reconnection events most likely induced a slipping reconnection in the yellow field lines (Figure \ref{nlfff}) carrying energy from the precursor location to the northern leg of the hot channel leading to its destabilisation. The entire mechanism suggested here is analogous to the domino effect that involve a sequence of destabilising processes that eventually cause a large-scale eruption \citep{Zuccarello2009}.

The AR experienced significant flux emergence for $\approx$10 hrs and cancellation thereafter for $\approx$2 hrs prior to the onset of the flare (Figures \ref{hmi_lightcurve}c and d). However, GOES SXR and AIA EUV light curves suggests absence of flaring activity in the AR during this prolonged period (\textit{cf.} Figures \ref{hmi_lightcurve}c, d and e) which signifies continuous storage of magnetic energy into the flaring environment without significant dissipation by reconnection events. Therefore, the emergence of magnetic flux in the trailing sunspot region and the subsequent phase of its decay prior to the onset of the flare possibly resulted in the build up of the MFR along the PIL. This is supportive of the flux cancellation model proposed by \citet{van1989} which states that flux cancellation at the PIL of a sheared magnetic arcade leads to the formation of MFRs. The build up of the MFR over the PIL in the trailing sunspot group of AR 12371 prior to the M-class flare on 21 June 2015 was observationally inferred by continuous brightening up of the hot channel as observed in AIA 94 \AA\ channel images (Figure \ref{fl94}a--c). Our analysis suggests that the hot channel continued to acquire the magnetic field and stress in response to the flux cancellation for an elongated period of time during the pre-flare phase. As a result, the MFR had already reached to a meta-stable state prior to the onset of the flare which allowed immediate eruption of it once triggered by the slipping reconnection. This type triggering can be explained by the ``tether-weakening'' model proposed by \citet{Moore1992}. According to this model, a meta-stable flux rope can be triggered by activities occurring adjacent to it, \textit{i.e.} off the main PIL \citep[see \textit{e.g.,}][]{Sterling2007,Yang2019}.

It is very interesting to note that, the flare originated from the same AR on the next day, (\textit{i.e.} 22 June 2015) proceeded with the formation and eruption of a hot channel from the trailing sunspot group \citep[see][]{Cheng2016b, WangH2017, Awasthi2018}, similar to the AIA observations of the M-class flare on 21 June, being investigated here. Also, in both the cases, the flux rope eruptions resulted into halo CMEs. However, the underlying flux rope structure of the hot channel bore differences between the events of 21 June and 22 June. By employing NLFFF extrapolation technique, \citet{Awasthi2018} observed multiple braided flux ropes with different degrees of coherency over the PIL during the pre-flare phase, which were separated in height. They also found evidence of small-scale reconnection events among the different flux rope branches which resulted into further braiding among the flux rope threads. While \citet{Awasthi2018} found the internal structure of the MFR on 22 June 2015 to be complex braided magnetic field, our result suggests coherent twisted field as the structure of the MFR. Further, there seems to have differences in the time-sequence and activities associated with the early stages of the MFR activation. Using high resolution observations of the precursor phase of the M6.5 flare on 22 June 2015 from the 1.6-m \textit{New Solar Telescope}, \citet{WangH2017} found two episodes of small-scale precursor brightening at the magnetic channel prior to the large-scale eruption of the MFR. Based on these observations, they concluded that low-atmospheric small-scale energy release events possibly triggered the eruption which supports the model proposed by \citet{Kusano2012}.

During the build up of the flux rope in the trailing sunspot prior to the M-class flare on 21 June, magnetic free energy stored in the AR increased significantly for a period of $\approx$5 hours which drastically reduced during the flare  (Figure \ref{imfe}). The M-class flare occurring on the next day from the same AR was subjected to similar type of magnetic energy evolution. These results clearly refer to a direct correlation between accumulation of free energy in the AR and build up of MFR with excess free energy in the form of their twisted (or, braided) magnetic structures.

In summary, the AR NOAA 12371 went through an elaborate phase of flux enhancement followed by a duration of significant flux cancellation which led to build up of an MFR along the PIL in the trailing sunspot group. The AR was associated with highly dynamical features including photospheric motions (\textit{i.e.} MMFs) and rotation which led to formation of localized regions of high photospheric current densities. Two well identified precursor events preceded the flux rope activation. The precursor region, spatially separated from the location of the MFR, exhibited strong photospheric longitudinal currents of opposite polarities in very close proximity and was connected directly to the MFR through magnetic loops. More importantly, we found evidence for slipping reconnection from the precursor region to the flux rope activation site which eventually destabilised the quasi-evolving MFR, resulting a halo CME and M-class flaring activities. Our study especially addresses the build-up phase of the MFR and the role of precursor activities toward driving the eruption. In future, we aim to study the slipping reconnection in the context of physical processes during the pre-flare and early flare evolution.

\begin{acks}
The authors would like to thank the SDO and RHESSI teams for their open data policy. SDO is NASA's mission under the Living With a Star (LWS) program. RHESSI was the sixth mission in the SMall EXplorer (SMEX) program of NASA. The authors are also thankful to Dr. Thomas Wiegelmann for providing the NLFFF code. AP acknowledges partial support of NASA grant 80NSSC17K0016 and NSF award AGS-1650854. We also acknowledge the constructive
comments and useful suggestions of the anonymous referee, which improved the presentation and scientific content of the article.\\
\\
\textbf{Disclosure of Potential Conflicts of Interest} ~~The authors declare that they have no conflict of interest.
\end{acks}

\end{article} 
\end{document}